\documentclass{article}

\usepackage{PRIMEarxiv}

\usepackage{hyperref}

\usepackage{graphicx}
\graphicspath{ {./plots/} }
\usepackage[version=3]{mhchem}
\usepackage[utf8]{inputenc}
\usepackage{amsmath}
\usepackage{multirow}
\usepackage[table,xcdraw]{xcolor}
\usepackage{amssymb}
\usepackage{amsfonts}
\usepackage{amstext}
\usepackage{amsthm}
\usepackage{mathrsfs}
\usepackage{mathtools}
\usepackage{caption}
\usepackage[linesnumbered,ruled,vlined]{algorithm2e}
\usepackage{mathtools}
\usepackage{float}
\usepackage{setspace}

\newtheorem{rem}{Remark}
\newtheorem{prop}{Proposition}
\newtheorem{them}{Theorem}
\title{A Reinforcement Learning-based Economic Model Predictive Control Framework for Autonomous Operation of Chemical Reactors}

\author{
  Khalid Alhazmi\\
  KAUST \\
  Thuwal, SA\\

  \texttt{khalid.alhazmi@kaust.edu.sa} \\
   \And
  Fahad Albalawi \\
  KAUST \\
  Thuwal, SA\\
  \texttt{fahad.albalawi@kaust.edu.sa} \\
  
  \And
  S. Mani Sarathy \\
  KAUST\\
  Thuwal, SA \\
  \texttt{mani.sarathy@kaust.edu.sa} \\
}

\begin{document}
\maketitle
\begin{abstract}
    Economic model predictive control (EMPC) is a promising methodology for optimal operation of dynamical processes that has been shown to improve process economics considerably. However, EMPC performance relies heavily on the accuracy of the process model used. As an alternative to model-based control strategies, reinforcement learning (RL) has been investigated as a model-free control methodology, but issues regarding its safety and stability remain an open research challenge. This work presents a novel framework for integrating EMPC and RL for online model parameter estimation of a class of nonlinear systems. In this framework, EMPC optimally operates the closed loop system while maintaining closed loop stability and recursive feasibility. At the same time, to optimize the process, the RL agent continuously compares the measured state of the process with the model’s predictions (nominal states), and modifies model parameters accordingly. The major advantage of this framework is its simplicity; state-of-the-art RL algorithms and EMPC schemes can be employed with minimal modifications. The performance of the proposed framework is illustrated on a network of reactions with challenging dynamics and practical significance. This framework allows control, optimization, and model correction to be performed online and continuously, making autonomous reactor operation more attainable.
\end{abstract}


\section{Introduction}

Model-based control and optimization is the predominant paradigm in process systems engineering, however, model parameters change over the operation cycle of chemical processes due to various causes, such as catalyst deactivation, equipment aging, feedstock variability, and more. Future autonomous engineering systems will require models that adapt to the changing characteristics of the environment \cite{lamnabhi2017systems}. A cornerstone of the operation of chemical plants and processes is calculating the optimum operating conditions and maintaining them, despite the presence of measurement uncertainties and disturbances \cite{seborg2010process}.

In efficient chemical plant operation, it is typical to first calculate the optimal steady state operating condition via a real time optimization (RTO) layer based on a steady-state model of the process \cite{marlin1997real}. The optimal set-points are then passed to supervisory control layers usually equipped with model predictive control (MPC) schemes to track these optimal set-points \cite{backx2000integration,de2010real,adetola2010integration}. Despite the positive impact of this control architectures on chemical plants, there are two main challenges associated with this scheme: First, process conditions -such as varying feed rate and equipment condition- will lead to a deviation from the economically optimal steady-state calculated in the RTO layer. Second, an accurate model is needed in order to calculate the optimal operating conditions. 
Promising methodologies, such as economic model predictive control (EMPC), have been proposed to address the first challenge. EMPC combines both RTO and MPC in one layer, minimizing the delay between process changes and the calculation and implementation of the new optimal set-points. Unlike the quadratic objective function associated with tracking MPC, EMPC incorporates a general cost function that directly accounts for process economics such as process yield or production rate \cite{rawlings2012fundamentals,ellis2014tutorial}. But even with this important advancement, the performance of EMPC still relies on the accuracy of the process model used. 

A promising alternative to model-based control strategies is the model-free reinforcement learning (RL) strategy. The aim of RL is to find a sequence of control actions that will maximize a predefined reward function. The RL agent interacts with the environment by applying an action that changes the state of the environment; the environment then generates a reward for that action. Using this information, the agent can adjust the necessary action in the future \cite{sutton2018reinforcement}. Reinforcement learning has recently attracted attention due to its success in learning how to play complex games, such as the board game Go and multiplayer poker, without human knowledge or supervision \cite{silver2016mastering,silver2017mastering,brown2019superhuman}. Successful applications of RL have spanned several fields, such as robotics, natural language processing, and computer vision \cite{nguyen2020deep, li2016deep, mnih2013playing}.

The RL framework can easily be related to that of optimal control if the environment is considered as a system, the action as a control, the agent as a controller, and the reward as the stage cost to be optimized \cite{bertsekas2019reinforcement}. Given the similarity between the two -as well as recent advances in RL- employing RL for control applications is a natural step toward targetting control challenges for which traditional control schemes are inadequate \cite{recht2019tour}. Several RL strategies have been proposed for process control problems and the reader is referred to references \cite{govindhasamy2004reinforcement,shin2019reinforcement,bucsoniu2018reinforcement} for reviews related to the application of reinforcement learning for process control. Early implementation of reinforcement learning paradigms for chemical reactions was proposed by Hoskins and Himmelblau \cite{hoskins1992process}, when the authors applied neural network architecture to a nonlinear continuous-stirred tank reactor (CSTR) with a simple hypothetical reaction \cite{hoskins1992process}. More recently, Zhou et al.\cite{zhou2017optimizing} developed a RL scheme for optimizing microdroplet reactions; they reported promising results \cite{zhou2017optimizing}. However, closed-loop stability and process constraint enforcement are challenging within the scope of RL, which hinders their complete replacement of model-based control strategies \cite{bucsoniu2018reinforcement}. 

Considering the pros and cons of EMPC and RL for process control applications suggests that integrating the two techniques can be a promising approach to utilizing the advantages of both methods and mitigating the drawbacks of each. A number of proposed algorithms have recently emerged to achieve learning-based control paradigms \cite{anderson2007robust,aswani2013provably,koller2018learning,wu2019real}. Efforts to integrate EMPC (in particular), and RL have focused on addressing the safety concerns resulting from exploration in RL. For instance, one approach is to use MPC as a function approximator in the RL scheme \cite{gros2019data,zanon2020safe}. While this approach is interesting, significant effort in adopting RL algorithms is required; so far, this has been done for relatively simple algorithms, such as Q-learning \cite{zanon2019practical}. MPC algorithms are commonly used to control chemical production processes, such as ethylene oxide. The models used within the MPC for these applications include time-varying parameters that are either costly to experimentally validate or theoretically difficult to estimate using conventional observers like an extended Kalman filter (EKF) and adaptive observers. Therefore, a general framework of integrating EMPC and RL agent for a class of nonlinear systems that can continuously update nonlinear parameters of the system is required. 

Motivated by the above considerations, a framework is proposed here that combines EMPC and RL, termed as RL-based EMPC, to estimate time varying parameters of chemical processes that follows a class of nonlinear systems. While proving stability for a general class of nonlinear systems is still under investigation in the control literature as well as in the general AI literature, the scope of the proposed work is not to utilize a RL agent to compute control actions that are responsible for both optimizing closed-loop performance and ensuring the closed-loop stability.  Instead, this approach uses RL to estimate unknown but bounded model parameters. The main objective of using the RL agent is to estimate the values of these unknown model parameters where the RL agent is restricted to learn the values between their corresponding upper and lower limits. The control actions that will be applied to the real system are then computed by economic MPC. When certain stabilizability and identifiability assumptions are fulfilled, practical stability and recursive feasibility for the proposed RL-based EMPC paradigm can be rigorously proven. To demonstrate the applicability of the proposed algorithm, a network of chemical reactions for ethylene oxide production is considered here.

\section{Preliminaries}
\subsection{Notation}
The transpose of the vector $x$ is represented by the symbol $x^T$. Symbols $||x||$ and $|x|$ signify the second norm of a vector $x$ and the absolute value of $x$, respectively. Vector $x$ represents observed states, while $\tilde{x}$ represents predicted states. The symbol $S\left(\Delta\right)$ denotes the family of piecewise constant functions with period $\Delta\geq0$. In the reward function, the notation $w(e(t)<\epsilon)$ represents a piecewise function that is equal to $w$ for all values of $e\left(t\right)$ in the subdomain $e\left(t\right)<\epsilon$, and zero otherwise, where $\epsilon$ is the error tolerance. The Greek letter $\theta$ symbolizes the tuning parameters of a nonlinear dynamic model, while the symbol $\phi$ denotes the set of parameters on which the reinforcement learning policies depend; in the case of neural network policies, the parameters are the weights and biases. The set subtraction between two sets $A_1$ and $A_2$ is signified by the symbol '/' (i.e., $A_1/A_2:= \{ x\in {R}^n: x\in A_1, x\notin A_2 \}$) 
\subsection{Class of nonlinear process systems}
The class of nonlinear process systems for which the proposed framework is applied is of the form: 
\begin{equation} \label{system}
    \dot x =f(x(t),\theta(t),u(t),d(t))
\end{equation}
where $x\left(t\right)\in\mathbb{R}^{n_x}$, $u\left(t\right)\in\mathbb{R}^{n_u}$, $\theta \left(t\right)\in\mathbb{R}^{n_\theta}$ and $d\left(t\right)\in\mathbb{R}^{n_d}$ are the state, control input, model parameters and disturbance vectors, respectively. The admissible controls are in the control region $U\subset\mathbb{R}$, where $U{\coloneqq}\left\{u\in\mathbb{R}^{n_u}\colon u^L\le u\left(t\right)\le u^U\right\}$, and $u^U$ and $u^L$ are the upper and lower limits, respectively. The disturbance vector $d\left(t\right)$ and the model parameters $\theta(t)$ are bounded in the following sets\\
$\mathbb{D}{\coloneqq}\left\{d\in\ \mathbb{R}^{n_d}\colon |d|\le\delta,\ \delta>0\right\}$ and $\mathbb{P}{\coloneqq}\left\{\theta\in\ \mathbb{R}^{n_\theta}\colon \theta_L\leq \theta\leq \theta_U\right\},$ respectively. It is assumed that the vector function $f$ is smooth and the origin is the equilibrium point for the unforced system (i.e., $0\equiv f(0,0,0,0)$). 
\subsection{Stabilizability Assumption}\label{lyapun}
Along the same lines of the stabilizability assumption in \cite{albalawi2016feedback}, it is assumed that there exists a Lyapunov-based controller $h(x)$, so that the origin of the nominal system (Eq. \ref{system} with $d(t)\equiv0$) is asymptotically stable with $h_i(x)\in U_i$, $i=1,\ldots,n_u$, inside a given stability region $\Omega_\rho$. Additionally, the existence \cite{khalil2002nonlinear,massera1956contributions} of a continuously differentiable Lyapunov function $V(x)$ is asserted for the nominal closed-loop system and a class $\mathcal{K}$ functions $\alpha_i(\cdot),\;i=1,2,3,4$ such that the following inequalities hold:
\begin{equation}\label{eqn:converse11}
	\begin{array}{c}
		\alpha_1(||x||) \leq V(x) \leq \alpha_2(||x||) \vspace{1mm}\\
		{\displaystyle\frac{\partial V(x)}{\partial x}}f(x,h_1(x),\ldots,h_m(x),0) \leq -\alpha_3(||x||) \vspace{1mm}\\
		{\left\Vert\displaystyle\frac{\partial V(x)}{\partial x}\right\Vert} \leq
		\alpha_4(|x|)\vspace{1mm}\\
		h_{i}(x) \in U_i,\;i=1,\ldots,m
	\end{array}
\end{equation}
for all $x \in D \subseteq R^{n}$ where $D$ is an open
neighborhood of the origin. We define a level set of the Lyapunov function within which $\dot{V}$ is negative as the stability region $\Omega_{\rho}$ of the process of Eq. \ref{system} under $h(x)$ (where $\Omega_{\rho} \subseteq D$; see, for example \cite{christofides2005control,kokotovic2001constructive} for results on the design of stabilizing control laws). 

Let $x$ be inside the stability region $\Omega_{\rho}$, $u_i \in U_i$, and $d \in \mathbb{D}$. The continuity of $x$, the local Lipschitz property of $f$, and the continuous differentiability of $V(x)$ imply that there exist positive constants $M$, $L_x$, $L_d$, $L_\theta$, $L^{*}_x$, $L_\theta^{*}$ and $L^{*}_d$, such that the following inequalities hold:
\begin{equation}\label{eqn:lip111}
	\left\Vert f(x(t),\theta(t),u(t),d(t))\right\Vert \leq M
\end{equation}
\begin{equation}\label{eqn:clfcont111}
	\begin{aligned}
		\left\Vert f(x,\theta,u,d)-f(x^{*},\theta^{*},u,0)\right\Vert\leq L_x||x-x^{*}||+L_\theta||\theta-\theta^{*}||+L_d||d||
	\end{aligned}
\end{equation}
\begin{equation}\label{eqn:clfcontt111}
	\begin{aligned}	
		\left\Vert \dfrac{\partial V(x)}{\partial x}f(x,\theta,u,d)-\dfrac{\partial V(x^{*})}{\partial x}f(x^{*},\theta^{*},u,0)\right\Vert\leq L^{*}_x||x-x^{*}||+L^{*}_\theta||\theta-\theta^{*}||+L^{*}_d||d||	
	\end{aligned}
\end{equation}
\noindent for all $x,x^{*}\in\Omega_{\rho}$, $u_i\in U_i$, $i=1,\ldots,m$, and $d\in \mathbb{D}$. 

\subsection{Lyapunov-Based Economic Model Predictive Control (LEMPC)}
Lyapunov-based economic model predictive control (LEMPC) is a form of EMPC formulation in which Lyapunov-based constraints were incorporated to ensure closed-loop stability and recursive feasibility. The mathematical formulation of Lyapunov-based EMPC is as follows \cite{heidarinejad2012economic}:
\begin{subequations}
	\label{eq:ecosysid:lempc11}
	\begin{align}
		\min_{u \in S(\Delta)} \quad & \int_{t_k}^{t_{k+N}} L_e(\tilde{x}(\tau), u(\tau)) ~d\tau \label{eq:ecosysid:lempc:cost11} \\
		\text{s.t.} ~\quad & \dot{\tilde{x}}(t) = f(\tilde{x}(t),\theta(t),u(t),0) \label{eq:ecosysid:lempc:model11} \\
		& \tilde{x}(t_k) = x(t_k) \label{eq:ecosysid:lempc:IC11} \\
		& u_i(t) \in U_i,~i = 1,\ldots,m, ~\forall ~t \in [t_k, t_{k+N}) \label{eq:ecosysid:lempc:input11} \\
		& V(\tilde{x}(t)) \leq \rho_e, ~\forall ~t \in [t_k, t_{k+N}) \nonumber \\
		& \qquad \text{if} ~x(t_k) \in \Omega_{\rho_e} \label{eq:ecosysid:lempc:mode111} \\
		& \frac{\partial V(x(t_k))}{\partial x} f(x(t_k), \theta(t_k), u(t_k), 0) \nonumber \\
		& \quad \leq \frac{\partial V(x(t_k))}{\partial x} f(x(t_k), \theta(t_k), h(x(t_k)), 0) \nonumber \\
		& \qquad \text{if} ~x(t_k) \notin \Omega_{\rho_e} \label{eq:ecosysid:lempc:mode211}
	\end{align}
\end{subequations}
where $\tilde x(t)$ is the predicted state, $u(t)$ is the control input, and $N$ is the prediction horizon with sampling periods of length $\Delta$. Eqn. \ref{eq:ecosysid:lempc:cost11} is the objective function of the LEMPC, and $L_e(\tilde{x},u)$ is the cost incurred at each stage. The predicted state is determined by the dynamic system of Eqn. \ref{eq:ecosysid:lempc:model11}, and the initial value of this optimization problem derives from a state measurement of the process at time $t_k$ (Eqn. \ref{eq:ecosysid:lempc:IC11}). The control input is restricted in the set shown in Eqn. \ref{eq:ecosysid:lempc:input11}, where admissible controls are defined in the region $ U_i$.

The Lyapunov-based constraints in Eqn. \ref{eq:ecosysid:lempc:mode111} and Eqn. \ref{eq:ecosysid:lempc:mode211} demonstrates two modes of operating the LEMPC. Mode 1 is activated when the present state is within the stability region $\Omega_{\rho_e}$, which is a subset of $\Omega_{\rho}$. On the other hand, Mode 2 is activated when the present state is within $\Omega_{\rho}$ but outside $\Omega_{\rho_e}$. The implementation of this constraint guarantees that the process under this Lyapunov-based EMPC is always within the stability region $\Omega_{\rho}$, despite the presence of disturbances and model mismatch. 

\section{Reinforcement Learning-based LEMPC}
\subsection{Reinforcement learning}
The objective of model-free reinforcement learning (RL) is to learn a policy ($\pi_\phi$) that maximizes the expected total discounted reward, $R$
\begin{gather}
    \max_{\pi_\phi} ~ {\mathbb{E}~\left[\sum_{k=0}^{\infty}{\gamma^k R(x(t_k),{\tilde{x}}(t_k)})\right]}\\
    \text{s.t.} ~~ ~~\theta(t_k)=\pi_\phi(\tau(t_k))\\
    \theta^L\le \theta_{t_k}\le \theta^U
\end{gather}
where $\pi_\phi$ is a control policy parametrized by $\phi$, and $\gamma$ is a discount factor. The symbol $\tau(t_k)$ denotes the predicted state sequence and the measured state sequence up to time $t_k$. An optimal policy can be learned in several ways \cite{arulkumaran2017brief}. The selected algorithm in this work is referred to as the deep deterministic policy gradient (DDPG) \cite{lillicrap2015continuous}. This algorithm is based on actor-critic methods, and it is suitable for process control applications as it allows for continuous action and state space. The architecture of the actor and critic neural networks is adopted from the work of Lillicrap et al. \cite{lillicrap2015continuous}, where each neural network has two hidden layers of 400 and 300 neurons, respectively.

The reward function, that is to be maximized, compares the measured plant states (i.e., real states) with the predicted states. A positive reward, $w_1$, is then given for a deviation within the accepted tolerance, $\epsilon$, and a negative reward, $w_2$, is given for a deviation that is greater than the accepted tolerance. The reward function is written as follows:
\begin{gather}
    R\left(x(t_k),\tilde{x}(t_k)\right)=\sum_{i=1}^{n_x}\left[w_1\left(e_i\left(t_k\right)<\epsilon\right)-w_2\left(e_i\left(t_k\right)>\epsilon\right)\right]\\
    e_i\left(t_k\right)=\frac{\left|x_i\left(t_k\right)-{\tilde{x}}_i\left(t_k\right)\right|}{x_i\left(t_k\right)}\ 
\end{gather}
where $e_i$ is the error in the prediction of the $i^{th}$ state; $w_1$ and $w_2$ are constants. The  training strategy of the RL agent for the nonlinear system of Eq. \ref{system} is depicted in Algorithm 1.  
\BlankLine

\begin{algorithm}[H]
    \SetAlgoLined
    \text{Initialize the neural networks according to the algorithm in ref \cite{lillicrap2015continuous}}\\
    \For{$episode\leftarrow 1$ \KwTo $F$}{ 
    \text{Reset the training environment to generate $\theta(t_0)$, $x(t_0)$}\\
        \For{$k\leftarrow 0$ \KwTo $T_{tr}$}{
            Solve the LEMPC of Eq. \ref{eq:ecosysid:lempc11} to obtain $u(t_{k})$\\
            Implement $u(t_{k})$\\
            Measure $x(t_{k+1})$ \\
            \text{Execute the inner For loop of the DDPG algorithm \cite{lillicrap2015continuous}}\\
            Compute $\theta(t_{k+1})=\pi_\phi( x(t_{k+1}),\tilde x(t_{k+1}))$\\
        }
    }
    \caption{RL Training Strategy}
\end{algorithm}
\vspace{35pt}
\textbf{A walk-through of Algorithm 1:}
\begin{itemize}

\item Step 1: In this work, the policy of the RL agent is chosen to be a neural network (NN) parametrized by $\phi$ (i.e., the weights of the neural network). In this step, the neural network parameters are randomly initialized.

\item Step 2: Before this step is illustrated, it is necessary to differentiate between an episode for the NN algorithm (also called a trajectory) and a step within the episode. An episode is a sequence of control actions and states, while a step is each update within the episode. In Algorithm 1, there are $F$ episodes and $T_{tr}$ updates within each episode. In this step, the counter for the $F$ episodes is increased by one. It is necessary to choose an $F$ large enough so that the RL agent has sufficient experience to learn a good policy, otherwise poor experimental results will be the outcome.

\item Step 3: At the beginning of each episode, the training environment is reset, when the tuning parameters, $\theta$, of the prediction model and the initial state of the plant are generated randomly within a specified range. The re-setting of the environment is essential for training the RL agent, otherwise the learned policy will be overfitted to a specific initial state. Note that the environment is not reset within one episode

\item Step 4: The counter for the length of the training period is updated by one step.  

\item Step 5: The LEMPC of Eq. \ref{eq:ecosysid:lempc11} is solved based on the current measurement $x(t_k)$ and parameters $\theta_{t_k}$, to obtain the control solution $u(t_k)$ 

\item Step 6: The control action $u(t_k)$ is applied to the system of Eq. \ref{system} in a sample-and-hold fashion. 

\item Step 7: When the control action is implemented, the updated process state $x(t_{k+1})$ can be measured. 

\item Step 8: In this step, the inner loop of the DDPG algorithm in ref \cite{lillicrap2015continuous} is followed. First, an action is selected by the algorithm (here, the action is $\theta$). Then, the action is executed, and the reward at time $t_{k+1}$ is calculated. Finally, the policy parameters, $\phi$, of RL agent are updated, generating a new policy.

\item Step 9: New model parameters are computed using the newly generated policy. The new parameters are then fed back to the LEMPC to be used in the next time step.

\end{itemize}

\subsection{RL-based LEMPC implementation strategy}
\begin{figure}[H]
\centering
\includegraphics[width=\textwidth]{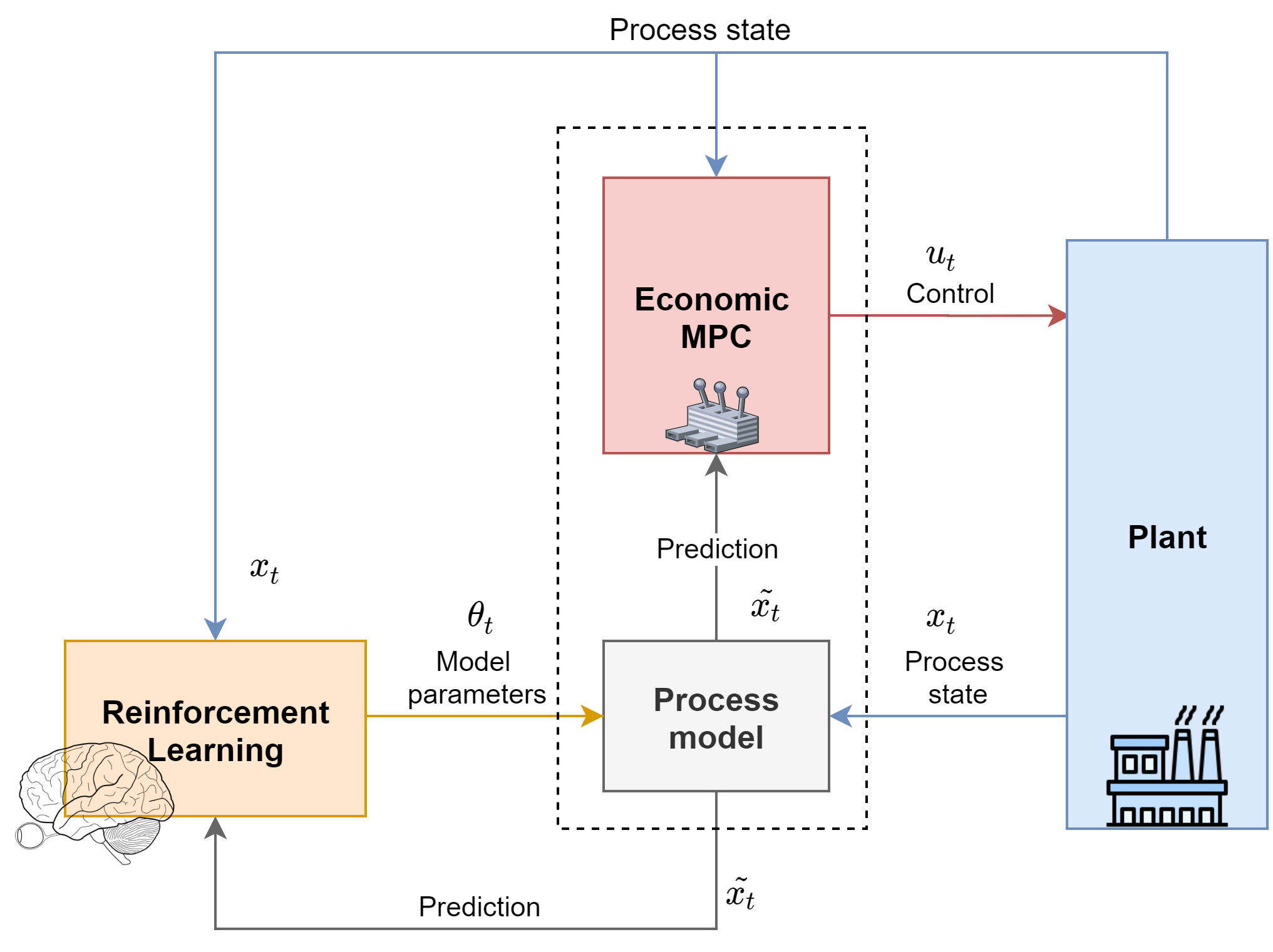}
\caption{Schematic representation of the integration of reinforcement learning with economic MPC}
\label{framework}
\end{figure}

The real-time interaction between the LEMPC of Eq. \ref{EMPC} and the RL agent along with their implementation strategy is illustrated in Figure \ref{framework} and Algorithm 2. The LEMPC perceives the current state $x(t_k)$ of the plant from the sensors and predicts the future state through the process model. The controller takes the subsequent appropriate actions $u(t_k)$ to optimize the closed-loop performance while meeting stability and input constraints. The RL agent compares the measured states with the predicted states and modifies the model parameters $\theta(t_k)$, accordingly. It should be noted that the main objective of the RL agent in this framework is to learn a policy that responds to plant-model mismatch caused by the continuous modification of the kinetic parameters.

\vspace{7pt}
\begin{algorithm}
    \SetAlgoLined
    \textbf{Import:~}$\pi_\phi$, $\theta(t_0)$, $x(t_0)$\\
    \vspace{11pt}
    \For{$k\leftarrow 0$ \KwTo $t_f$}{
        Solve the LEMPC of Eq. \ref{eq:ecosysid:lempc11} to obtain $u(t_{k})$\\
        \vspace{7pt}
        Implement $u(t_{k})$\\
        \vspace{7pt}
        Measure $x(t_{k+1})$ \\
        \vspace{7pt}
        Compute $\theta(t_{k+1})=\pi_\phi(x(t_{k+1}),\tilde x(t_{k+1}))$\\
        \vspace{7pt}
    }
    \caption{RL-based LEMPC}
\end{algorithm}

\vspace{17pt}
\begin{rem}
It is notable that the difference between Algorithm 1 and Algorithm 2 is that in Algorithm 2, the control policy $\pi_\phi$ is already learned and known from the training stage in Algorithm 1. Therefore, in the real-time implementation for the RL-based LEMPC, the appropriate value of the kinetic parameters $\theta$ can be computed immediately, once the measured state $x(t_k)$ and predicted state $\tilde x(t_k)$ are given.          
\end{rem}
\textbf{A walk-through of Algorithm 2:}
\begin{itemize}
\item Step 1: The trained RL agent $\pi_{\phi}$ is imported and the process states and kinetic parameters are initialized. 

\item Step 2: The counter for the operating period is initialized where $t_k$ denotes the current time step, and $t_f$ denotes the length of the operating period.

\item Step 3: The LEMPC of Eq. \ref{EMPC} is solved based on $x(t_k)$ and $\theta(t_k)$.

\item Step 4: In this step, the control action $u(t_k)$ is applied to the system of Eq. \ref{system} in a sample-and-hold fashion.   

\item Step 5: The control action is implemented, and the updated process state $x(t_{k+1})$ can be measured.  

\item Step 6: The model parameters are updated via $\pi_\phi$, $x(t_{k+1})$ and $\tilde x(t_{k+1})$. The new parameters are then fed back to the LEMPC to be used in the next time step. 

\end{itemize}

\section{Closed-loop Stability Analysis of RL-based EMPC}
This section provides the closed-loop stability and the recursive feasibility analysis of the proposed RL-based EMPC framework. First, Proposition \ref{prop11} is restated from \cite{heidarinejad2012economic}. Two propositions are then cited, with their detailed proofs. Finally, the main result of this note is established in Theorem 1.

\begin{prop} \label{prop11}
Consider the following two systems:
\begin{equation}
\dot x_n =f(x_n(t),\theta_n(t),u(t),d(t)) 
\end{equation}

\begin{equation}
\dot x_d =f(x_d(t),\theta_d(t),u(t),0)
\end{equation}
with initial states $x_n(t_0)$ and $x_d(t_0) \in \Omega_\rho$. There exists a class $\mathcal{K}$ function $f_d(.)$ so that 

\begin{equation}
||x_n(t)-x_d(t)||\leq f_d(t-t_0)
\end{equation}
for all $x_n(t), x_d(t) \in \Omega_\rho$, $\theta_n(t), \theta_d(t) \in \mathbb{P}$ and $d(t) \in \mathbb{D}$ with

\begin{equation}
f_d(\tau) = \frac{L_d\delta}{L_x L_\theta} (e^{L_xL_\theta \tau}-1)
\end{equation}

\end{prop}

In Proposition 1, an upper bound on the deviation between the state trajectory obtained from the nominal model and the state trajectory obtained from the disturbed system is derived when the same control input trajectories are applied. Inspired by the work in \cite{wu2020post}, Proposition 2 provides an upper bound on the deviation between the actual state and the estimated state obtained from the RL agent in Algorithm 1.

\begin{prop} \label{prop1}
Consider the real state $x(t)$ of the nonlinear system of Eq. \ref{system} and the estimated state $\tilde x(t)$ which is computed based on the updated value of $\theta(t_k)$ from the trained RL-agent of Algorithm 1 with the initial condition $|x(t_0)-\tilde x(t_0)| \leq \beta$, where $\beta>0$. When $x(t)$ and $\tilde x(t)$ are inside $\Omega_\rho$ for all times, a positive constant $\nu$ exists, so that the following inequalities hold for all $x(t)$, $\tilde x(t)$ $\in \Omega_\rho$:
\begin{equation} \label{first_res}
    ||x(t_0)-\tilde x(t_0)|| \leq \beta e^{L_xt} 
\end{equation}
\begin{equation} \label{second_res}
    V(x)\leq V(\tilde x) + \alpha_4(\alpha_{1}^{-1}(\rho))||x-\tilde x||+\nu ||x-\tilde x||^2
\end{equation}

\end{prop}

\begin{proof}
To prove the first result (i.e., Eq. \ref{first_res}) of Proposition \ref{prop1}, it is necessary to first define the error state as the difference between the actual state and the estimated one based on the updated value of $\theta(t_k)$. The error state vector is defined as $x(t)-\tilde x(t)$. Using Eq. \ref{eqn:clfcont111}, the time derivative of the error vector for all values of $x(t)$, $\tilde x(t)$ within $\Omega_\rho$ and $u \in U$ is as follows: 
\begin{align}
||\dot e|| = ||f(x,u) - f(\tilde x,u)|| \leq L_x ||x - \tilde x|| = L_x ||e(t)|| 
\end{align}
Since the error between $x_0(t)$ and $\tilde x_0(t)$ is bounded by $\beta$ ((i.e. $||x_0(t)-\tilde x_0|| \leq \beta$), the upper bound of $||e(t)||$ can subsequently be derived for all $x(t)$, $\tilde x(t) \in \Omega_\rho$ as follows: 
\begin{align}
 ||e(t)||=||x(t)-\tilde x(t)||\leq \beta e^{L_xt}
\end{align}

The second result of Eq. \ref{second_res} can be proved using the Lyapunov inequalities result of Eq. \ref{eqn:converse11} and the Taylor series expansion of $V(x)$ around $\tilde x$ for all $x(t)$, $\tilde x(t)$ $\in \Omega_\rho$ as follows: 
\begin{align}
     V(x) \leq V(\tilde x) + \frac{\partial V(\tilde x)}{\partial x} ||x-\tilde x|| + \nu ||x-\tilde x||^2 \\ 
     V(\tilde x) + \alpha_4(\alpha_{1}^{-1}(\rho)) ||x-\tilde x|| + \nu ||x-\tilde x||^2
\end{align}
\end{proof}

Also inspired by \cite{wu2020post}, Proposition \ref{prop3} demonstrates the fact that the nonlinear system of Eq. \ref{system} can be rendered negative for all times, so that the actual state $x$ can be driven towards the origin under implementation of the stabilizing controller $h(\tilde x)$, which utilizes the estimated state, derived from the updated parameters $\theta(t_k)$ computed by the RL agent of Algorithm 1.

\begin{prop} \label{prop3}
Consider the nonlinear system of Eq.\ref{system} under the receding horizon application of the stabilizing controller $u = h(\tilde x) \in U$ based on the estimated state $\tilde x$, derived from the updated parameters $\theta(t_k)$ computed by the RL agent of Algorithm 1, which satisfies $||x-\tilde x|| \leq \beta$. Let $\epsilon_s> 0,\Delta > 0$ and $\rho > \rho_s > 0$ satisfy
\begin{align} \label{condition}
 -\alpha_3(\alpha_{2}^{-1}(\rho_s)) + L^{*}_{x}(\beta + M \Delta) \leq \epsilon_s
\end{align}
Then, $\dot V(x) \leq - \epsilon_s$ is true $\forall~x(t_k)\in \Omega_{\rho}/\Omega_{\rho_s}$ 
\end{prop}

\begin{proof}
The proof of this proposition follows the same path as the proof of Proposition 4 in \cite{wu2019machine}, with only one modification. The modification is needed to take the estimation error $\beta$ into account in the Lie derivative of the Lyapunov function (i.e., $\dot V(\tilde x)$) as follows:  
\begin{align} 
    & \dot V(x(t_k)) = \frac{\partial V(x(t_k))}{\partial x} f(x(t_k),h(\tilde x(t_k))) \\\nonumber
    & =\frac{\partial V(\tilde x(t_k))}{\partial x} f(\tilde x(t_k),h(\tilde x(t_k))) ~+ \\\nonumber
    &\frac{\partial V(x(t_k))}{\partial x} f(x(t_k),h(\tilde x(t_k))) - \frac{\partial V(\tilde x(t_k))}{\partial x} f(\tilde x(t_k),h(\tilde x(t_k)))
\end{align}
Furthermore, the following inequalities can be derived using Eq. \ref{eqn:clfcontt111}, Eq. \ref{eqn:converse11} and the Lipschitz condition of Eq. \ref{eqn:lip111} as follows:
\begin{equation*}
\dot V(x(t_k)) \leq - \alpha_3(\alpha_{2}^{-1}(\rho_s)) + L^{*}_{x}||x(t_k)-\tilde x(t_k)|| 
\leq - \alpha_3(\alpha_{2}^{-1}(\rho_s)) + L^{*}_{x}\beta 
\end{equation*}

As a result, $\dot V(x(t_k))\leq - \epsilon_s$ can similarly be proved by accounting for the effect of the receding horizon implementation of the control action on the nonlinear system of Eq. \ref{system}, following the same path as the proof in \cite{wu2019machine} when the inequality of Eq. \ref{condition} is fulfilled.
\end{proof}

The main result of this note is described in the following theorem which demonstrates the fact that $\Omega_\rho$ can be made forward invariant set under the RL-based LEMPC (described in Algorithm 2) for the nonlinear system of Eq. \ref{system} when the set $\Omega_{\rho_e}$ is carefully chosen for the closed-loop system under the LEMPC of Eq. \ref{eq:ecosysid:lempc11}    
\begin{them} \label{ththth}
Consider the nonlinear system of Eq. \ref{system} under the receding horizon implementation of the RL-based EMPC of Algorithm 2. Let $\Delta>0$ and $\rho>\rho_e>\rho_s>0$ satisfy the following inequality:
\begin{equation}
\rho_e \leq \rho - \alpha_4(\alpha_{1}^{-1}(\rho))\beta e^{L_x \Delta} - \nu (\beta e^{L_x \Delta})^2   
\end{equation}
If $||\tilde x - x|| \leq \beta$ for all times, then, the real state of the nonlinear system of Eq. \ref{system} under the RL-based LEMPC of Algorithm 2 is guaranteed to stay inside the stability region $\Omega_\rho$, for all $t\geq 0$ for any $x_0 \in \Omega_\rho$.    
\end{them}

\begin{proof}
Like the results of Proposition \ref{prop1}, the subset of the stability region $\Omega_{\rho_e}$ is chosen to account for the estimation error between the actual state $x(t_k)$ and the estimated state $\tilde x(t_k)$ using the updated parameters $\theta(t_k)$ from the RL agent of Algorithm 1. In addition, the solution $u(t)=h(x(t))$ where $t\in[t_k,t_{k+1}]$, $k=0,1,..,N-1$ is a feasible solution to the RL-based EMPC of Eq. 6 because $u(t)=h(x(t))$ satisfies the input constraints of Eq. 6d by the definition of the Lyapunov-based controller $h(x)$. It also satisfies the stability constraints of Eq. 6e and Eq. 6f, due to the stability properties of the nonlinear controller $h(x(t))$. The detailed proof of Theorem \ref{ththth} is analogous to the proof of Theorem 2 in \cite{heidarinejad2012economic}, and is excluded here for brevity.         
\end{proof}
\vspace{10 pt}

\begin{rem}
To ensure asymptotic stability of the nonlinear system of Eq. \ref{system} under the RL-based LEMPC of Algorithm 2, the stability constraint of Eq. \ref{eq:ecosysid:lempc:mode211} can be enforced for all times which will eventually drive the closed-loop state to a small neighborhood around the origin, due to the stability properties of the Lyapunov-based controller $h(\tilde x(t_k))$. Nevertheless, activating this constraint for all times will significantly impact the process economics because of the fact that the operating region at which the closed-loop system can maximize process economics has been reduced from $\Omega_{\rho_e}$ to $\Omega_{\rho_s}$.         
\end{rem}

\section{Application}
\subsection{Reactor process modeling}
The RL-based EMPC framework is applied to the catalytic oxidation of ethylene to ethylene oxide in a continuous stirred-tank reactor (CSTR). This reaction resembles many industrial processes in which a metal-based heterogeneous catalyst is used for a complex gas-solid reaction. The reaction scheme consists of the following chain of reactions:
\begin{gather}
\ce{ C2H4 + 1/2 O2 ->[\text{$r_1$}] C2H4O } \label{rxn1}\\
\ce{ C2H4 + 3 O2 ->[\text{$r_2$}] 2 CO2 + 2 H2O } \label{rxn2}\\
\ce{ C2H4O + 5/2 O2 ->[\text{$r_3$}] 2 CO2 + 2 H2O \label{rxn3} }   
\end{gather}
The reaction rates are adopted from the work of Alfani and Carberry \cite{alfani1970exploratory} as reported by Ozgulcsen et al. \cite{ozgulcsen1992numerical}
\begin{align}
r_1=k_1\ exp\left(\frac{-E_1}{RT}\right)p_E^{0.5} \label{r1}\\ 
r_2=k_2\ exp\left(\frac{-E_2}{RT}\right)p_E^{0.25}\label{r2}\\
r_3=k_3\ exp\left(\frac{-E_3}{RT}\right)p_{EO}^{0.5} \label{r3}
\end{align}
where $r_i$, $E_i$, and $k_i$ are, respectively, the reaction rate, activation energy, and pre-exponential factor for the $i^{th}$ reaction. $R$ is the universal gas law constant, and $T$ is reactor temperature. The partial pressure of ethylene and ethylene oxide is denoted by $p_E$ and $p_{EO}$, respectively. The following ordinary differential equations (ODE’s) show the design equations of a non-adiabatic CSTR:
\begin{subequations} \label{cstr1}
    \begin{align}
        &\frac{d\rho}{dt^\prime}=\frac{Q_f}{V} \left(\rho_f-\frac{T}{T_f} \rho\right)\\
        &\frac{dC_E}{dt^\prime}=\frac{Q_f}{V} \left(C_{E,f}-\frac{T}{T_f} C_E \right)-\frac{w}{V} \left(r_1+r_2\right)\\
        &\frac{dC_{EO}}{dt^\prime}=\frac{Q_f T}{V T_f}C_{EO}+ \frac{w}{V} (r_1-r_3)\\
        &\frac {dT}{dt^\prime}=\frac{Q_f\rho_f}{V\rho} (T-T_f) +w \frac{(-\Delta H_1) r_1+(-\Delta H_2) r_2+(-\Delta H_3)r_3}{V\rho C_p}\\ \nonumber
        &~~~~~\quad -\frac{hA}{V\rho C_p} \left(T-T_c\right) \label{cstr2} 
    \end{align}
\end{subequations}
where the reaction parameters are given in Table 1. Note that $(\cdot)_f$ and $(\cdot)_c$ denote the feed and coolant conditions, respectively.

\begin{table}[H]
\centering
\caption{List of parameters}
\begin{tabular}{cl}
\hline
Symbol & Name of parameter \\
\hline
$\rho$       & Density of gas mixture                   \\
Q          & Volumetric flow rate                       \\
V          & Reactor volume                             \\
$C_E$        & Effluent concentration of ethylene       \\
$C_{EO}$     & Effluent concentration of ethylene oxide \\
w          & Weight of catalyst                         \\
$\Delta H_i$ & Heat of reaction i                       \\
$C_p$        & Heat capacity of gas mixture             \\
h          & Heat transfer coefficient                  \\
A          & Heat transfer area                      
\end{tabular}
\end{table}

Ozgulcsen et al. \cite{ozgulcsen1992numerical} describe the above CSTR system (Eqs \ref{cstr1}-\ref{cstr2}) in the following dimensionless form:

\begin{subequations}\label{ODEs}
    \begin{align}
        & \frac{dx_1}{dt}=u_1\left(1-x_1x_4\right)\\
        & \frac{dx_2}{dt}=u_1\left(u_2-x_2x_4\right)-A_1\theta_4\exp{\left(\frac{\gamma_1\theta_1}{x_4}\right)}\left(x_2x_4\right)^{0.5}\\ \nonumber
        & ~~~~~~~~~ -A_2\theta_5\exp{\left(\frac{\gamma_2\theta_2}{x_4}\right)}\left(x_2x_4\right)^{0.25}\\
        & \frac{dx_3}{dt}=-u_1x_3x_4+A_1\theta_4\exp{\left(\frac{\gamma_1\theta_1}{x_4}\right)}\left(x_2x_4\right)^{0.5}\\ \nonumber
        & ~~~~~~~~~  -A_3\theta_6\exp{\left(\frac{\gamma_3\theta_3}{x_4}\right)}\left(x_3x_4\right)^{0.5}\\
        & \frac{dx_4}{dt}=\frac{u_1}{x_1}\left(1-x_4\right)+\frac{B_1}{x_1}\exp{\left(\frac{\gamma_1}{x_4}\right)}\left(x_2x_4\right)^{0.5} +\frac{B_2}{x_1}\exp{\left(\frac{\gamma_2}{x_4}\right)}\left(x_2x_4\right)^{0.25}\\ \nonumber 
        & ~~~~~~~~~ +\frac{B_3}{x_1}\exp{\left(\frac{\gamma_3}{x_4}\right)}\left(x_3x_4\right)^{0.5}+\frac{B_4}{x_1}\left(x_4-u_3\right)
    \end{align}
\end{subequations}
where the dimensionless constants are defined as:
\begin{gather*}
x_1=\frac{\rho}{\rho_f},\ x_2=\frac{C_E}{C_{ref}},\ x_3=\frac{C_{EO}}{C_{ref}}, x_4=\frac{T}{T_f}\\
t=\frac{Q_{ref}t^\prime}{V},\ u_1=\frac{Q_f}{Q_{ref}},u_2=\frac{C_{E,f}}{C_{ref}},\ u_3=\frac{T_c}{T_f}\\
\gamma_1=-\frac{E_1}{RT_f},\ \gamma_2=-\frac{E_2}{RT_f},\ \gamma_3=-\frac{E_3}{RT_f}\\
A_1=\frac{9.06k_1wT_f^{0.5}}{Q_{ref}C_{ref}^{0.5}},\ A_2=\frac{3.01k_2wT_f^{0.25}}{Q_{ref}C_{ref}^{0.75}},\ A_3=\frac{9.06k_3wT_f^{0.5}}{Q_{ref}C_{ref}^{0.5}}\\
B_1=\frac{9.06k_1(-\Delta H_1)wC_{ref}^{0.5}}{{\rho_fC_pT_f^{0.5}Q}_{ref}},\ B_2=\frac{3.01k_2(-\Delta H_2)wC_{ref}^{0.25}}{{\rho_fC_pT_f^{0.75}Q}_{ref}}\\ B_3=\frac{9.06k_3(-\Delta H_3)wC_{ref}^{0.5}}{{\rho_fC_pT_f^{0.5}Q}_{ref}},\ B_4=\frac{hA}{\rho_fC_pQ_{ref}}
\end{gather*}
and their values are listed in Table 2. The unknown model parameters $\theta_1$ … $\theta_6$ are bounded within $[0.9,\ 1.1]$, and they represent tuning parameters for the pre-exponential factor and activation energy for each of the reactions.

\begin{table}[h!]
\centering
\caption{Values of dimensionless constants from Ozgulcsen et al. \cite{ozgulcsen1992numerical}}
\begin{tabular}{cccc}
\hline
Constant & Value & Constant & Value \\
\hline

$\gamma_1$ & -8.13  & $A_3$ & 2417.71    \\
$\gamma_2$ & -7.12  & $B_1$ & 7.32       \\
$\gamma_3$ & -11.07 & $B_2$ & 10.39      \\
$A_1$      & 92.80  & $B_3$ & 2170.57    \\
$A_2$      & 12.66  & $B_4$ & 7.02   
\end{tabular}
\end{table}

\subsection{EMPC formulation }
For the CSTR process of Eq. \ref{ODEs}, the control objective is to maximize the yield of ethylene oxide, given a certain ethylene feed rate and concentration, by manipulating the coolant temperature. We chose the yield of the desired product as an objective function because it is directly related to the plant’s profitability. To mimic real practice in chemical plants, the coolant temperature was chosen as the manipulated variable. The operator’s goal is to maximize the amount of desired product produced for whatever feed is available.

The CSTR process of Eq. \ref{ODEs} has four states: $x=\left[x_1 ~ x_2 ~ x_3 ~ x_4\right]^T$, as defined in the previous section. The reactor is initialized at the steady-state $x_s^T=\left[0.998 ~ 0.432 ~ 0.0292 ~ 1.002\right]$, which corresponds to the steady-state input $u_s^T = [0.2 ~ 0.5 ~ 1]$. The manipulated variable is the coolant temperature, $u_3$, which is constrained as follows:
$$0.6\le u_3\le 1.4\ \ $$
The other input variables, $u_1$ and $u_2$, are set to their steady-state values. The performance criterion for this system is the average yield of ethylene oxide as defined by the following relation:
$$Y(t_f)\ =\ \frac{\int_{0}^{t_f}{u_1\left(\tau\right)\ x_3\left(\tau\right)\ x_4\left(\tau\right)\ d\tau\ }}{\int_{0}^{t_f}{u_1\left(\tau\right)\ u_2\left(\tau\right)\ d\tau}}\ $$
Since the denominator is fixed, the stage cost to be optimized is the following equation of the CSTR of Eq. \ref{ODEs}:
$$l_e=\ u_1 \ x_3\ x_4\  $$
The following EMPC optimization problem is solved at each sampling time $t_k$ for the CSTR of Eq. \ref{ODEs}: 
\begin{subequations} \label{EMPC}
\begin{align}
\min_{u\ \in\ S\left(\Delta\right)}{\int_{t_k}^{t_k+N}{-\ u_1\left(\tau\right)\ {\tilde{x}}_3\left(\tau\right){\tilde{x}}_4\left(\tau\right)d\tau}}\\
\text{s.t.} ~~\dot{\tilde{x_k}}=f\left(\tilde{x}(t), \theta(t),u(t),0\right)\\
\tilde{x}\left(t_k\right)=x\left(t_k\right)\\
 0.6\le\ u_3\left(t\right)\le 1.4
 \end{align}
\end{subequations} 
where $f$ is a plant model parameterized by a vector $\theta=\left[\theta_1,\ \theta_2,\ \theta_3,\ \theta_4,\ \theta_5,\ \theta_6\right]$. Following Algorithm 2, the EMPC of Eq. \ref{EMPC} is first initialized at $x_s^T$ and $\theta(t_0)$. The first control $u(t_k)$ is then applied to the real plant in a sample-and-hold fashion (i.e., $\dot x(t)=f(x(t),\theta(t_k),u(t_k))~\forall t\in[t_k,t_{k+1}]$). Subsequently, the RL agent will compute an updated value for $\theta(t_{k+1})$ based on $\tilde x(t_{k+1}) - x(t_{k+1})$. After that, the updated value of $\theta$ along with the updated measured state $x(t_{k+1})$ will be fed back to the EMPC of Eq. \ref{EMPC} to compute a new control trajectory.        


\section{Results and Discussion}
\subsection{Training the RL agent}
A variable-step numerical differentiation formulas (NFDs)-based numerical ODE solver is used to simulate the dynamic model of Eq. \ref{ODEs}. The continuous nonlinear dynamic system of Eq. \ref{ODEs} is discretized using the Eurler method with a step size of 0.01. The nonlinear optimization problem of the LEMPC of Eq. \ref{EMPC} is solved using an nlmpc object in Simulink. The length of sampling period $\Delta = 1$, and the prediction horizon was ten. The RL agent is trained on the NVIDIA RTX 2060 Graphics Processing Unit (GPU). To allow computation on GPU's, MATLAB's Parallel Computing Toolbox is used to generate NVIDIA CUDA code. The number of updates within each episode, $T_{tr}$, is 400. Training required $F$ = 1468 episodes and lasted for 7.2 hours.

\begin{figure}[H]
\centering
\includegraphics[width=0.8\textwidth]{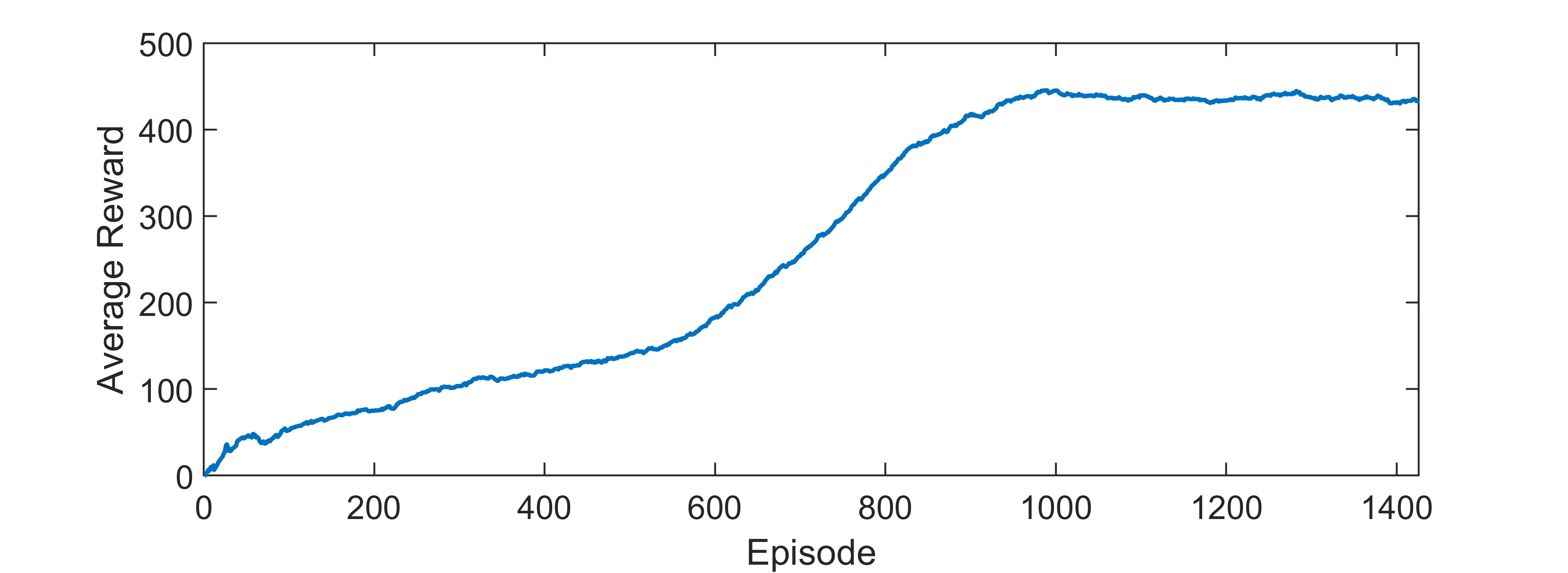}
\caption{Training RL agent on process described in Eqs. \ref{ODEs} with deep deterministic policy gradient (DDPG) algorithm and neural network architecture described by Lillicrap et al. \cite{lillicrap2015continuous}}
\label{averageReward}
\end{figure}

At each time step of the training, the kinetic parameters of the plant are randomly generated within the range $[0.9, 1.1]$. Being exposed to this experience, the RL agent attempts to learn an approximate policy. The learning behavior is shown in Figure \ref{averageReward}, where the average reward the agent receives increases with more learning episodes. It should be noted here that the RL agent is not trained on a particular scenario of how the plant and model mismatch occurs. That is, no correlations among the values of $\theta_1 ... \theta_6$ are assumed. The training is stopped when the change in the average reward with respect to training episodes begins to approach zero.

\subsection{Control and optimization performance}
To illustrate the utility of the proposed framework, a scenario was considered in which dynamic changes in the process (such as catalyst deactivation) could result in a mismatch with the process model. Catalyst deactivation leads to a decrease in catalytic activity and selectivity, and it occurs due to various causes, such as sintering, poisoning, and accumulation of deposits on the surface of the catalyst \cite{boskovic2004deactivation}. Several experimental studies examined the effect of catalyst deactivation on the kinetics of the oxidation of ethylene to ethylene oxide \cite{montrasi1983oxidation,borman1995experimental,boskovic2004deactivation}. For instance, Montrasi et al. \cite{montrasi1983oxidation} showed that the activation energy of ethylene formation with a deactivated catalyst is more than $30\%$ lower than that of a fresh catalyst.

To simulate the catalyst deactivation, the reaction rate of the desired reaction, Eq. \ref{r1}, was decreased by increasing the activation energy, $E_1$, and decreasing the rate constant, $k_1$. Conversely, the reaction rates of the undesired reactions, Eq. \ref{r2} and \ref{r3}, were increased by decreasing the activation energies and increasing the rate constants. In summary, the reaction parameters $E_1$, $k_2$, and $k_3$ were increased by $5\%$ over five steps, while the reaction parameters $k_1$, $E_2$, and $E_3$ were also decreased by $5\%$ over five steps as well.

\begin{figure}[H]
\centering
\includegraphics[width=0.8\textwidth]{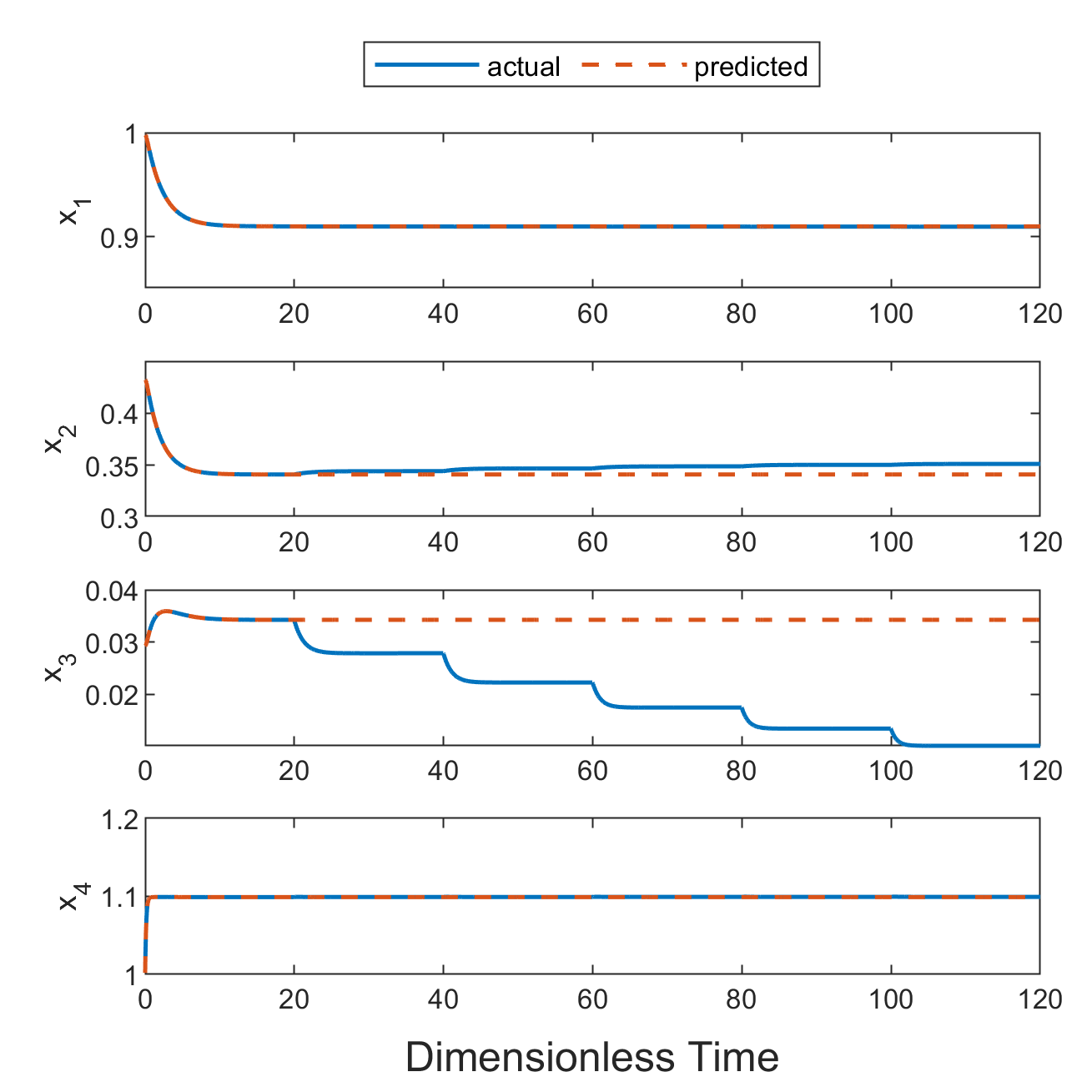}
\caption{State trajectories of process described by Eqs. \ref{ODEs} for EMPC without RL}
\label{x_base}
\end{figure}

Figure \ref{x_base} compares the values of the observed states of the plant to the values predicted by the plant model, which is used by the EMPC. Initially, the model was identical to the plant. However, after that the catalyst began to deactivate, the model prediction and the actual plant started to diverge considerably. As expected, the greatest deviation was in the ethylene and ethylene oxide concentrations, $x_2$ and $x_3$, respectively, since the gas density and reactor temperature were not strongly correlated with catalyst deactivation.

\begin{figure}[H]
\centering
\includegraphics[width=0.8\textwidth]{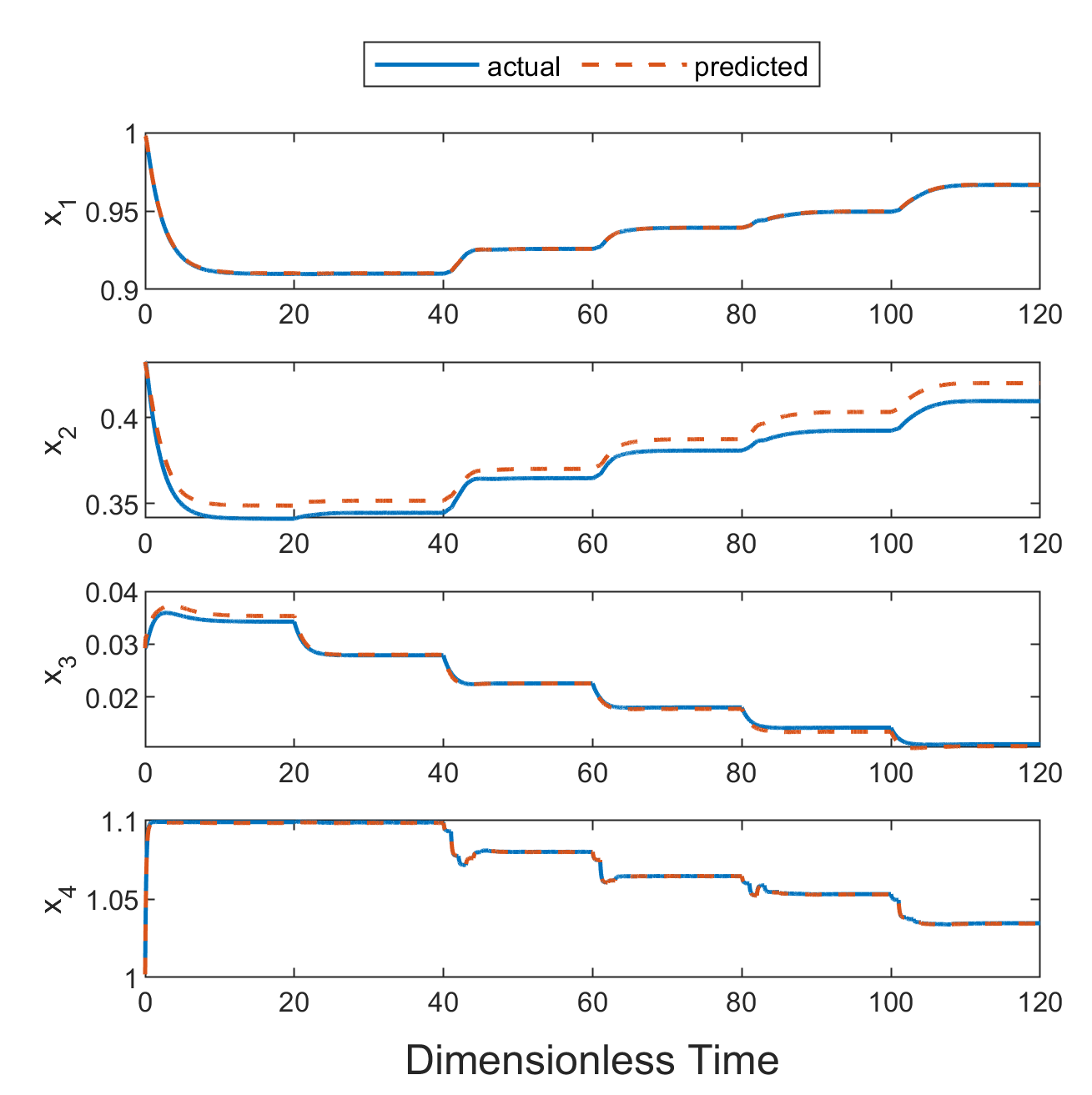}
\caption{State trajectories of process described by Eqs. \ref{ODEs} for EMPC-RL framework.}
\label{x}
\end{figure}

Conversely, when the EMPC-RL scheme was employed, deviation between the plant and the model predictions did not exceed $3\%$, as shown in Figure \ref{x}. As stated earlier, the objective of the RL agent was to learn a policy that responds to model mismatch by manipulating the kinetic parameters. Since this policy was represented by a neural network, it was not possible to justify a particular choice of kinetic parameters, which leads to a particular state prediction. For instance, in Figure \ref{x}, we notice that the prediction of state $x_3$ in the time steps 20 to 40 was more accurate than the prediction in time steps 0 to 20. This particular observation did not necessarily hold when a different RL algorithm or neural network architecture was used. All that can be said is that, regardless of what $\theta$ the RL agent chooses, the objective was to minimize the deviation between the observed and predicted states. 

Also, considering the first 20 time steps in Figure \ref{x}, a mismatch was noted in the prediction of states $x_2$ and $x_3$, despite the fact that the plant's kinetic parameters of the plant did not yet change (i.e. catalyst deactivation had not begun). This observation can be understood in light of the following two points. First, when RL was used along with EMPC, the kinetic parameters of the plant's model was solely determined by the RL agent. Second, the policy that the RL agent learned was not an optimal policy, but it was an approximation. This initial deviation, despite knowing the initial kinetic parameters a priori can be easily amended by forcing an initial action in the RL algorithm. However, this was not performed in this work, as it was important to emphasize this issue. 

\begin{figure}[H]
\centering
\includegraphics[width=\textwidth]{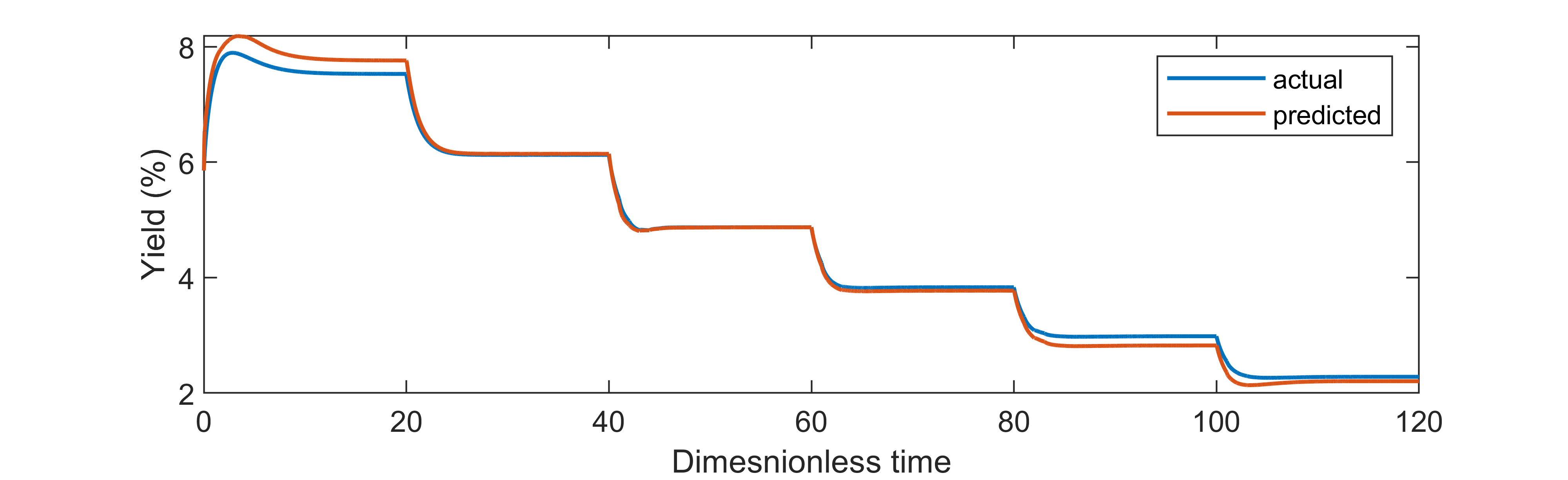}
\caption{Comparison of predicted yield with plant yield when proposed framework is used.}
\label{yield}
\end{figure}

Next, the yield predicted by the model was compared with the yield of the plant when the EMPC-RL scheme was employed. Figure \ref{yield} shows that the error in predicting the yield was less than $4\%$.

\begin{figure}[H]
\centering
\includegraphics[width=1\textwidth]{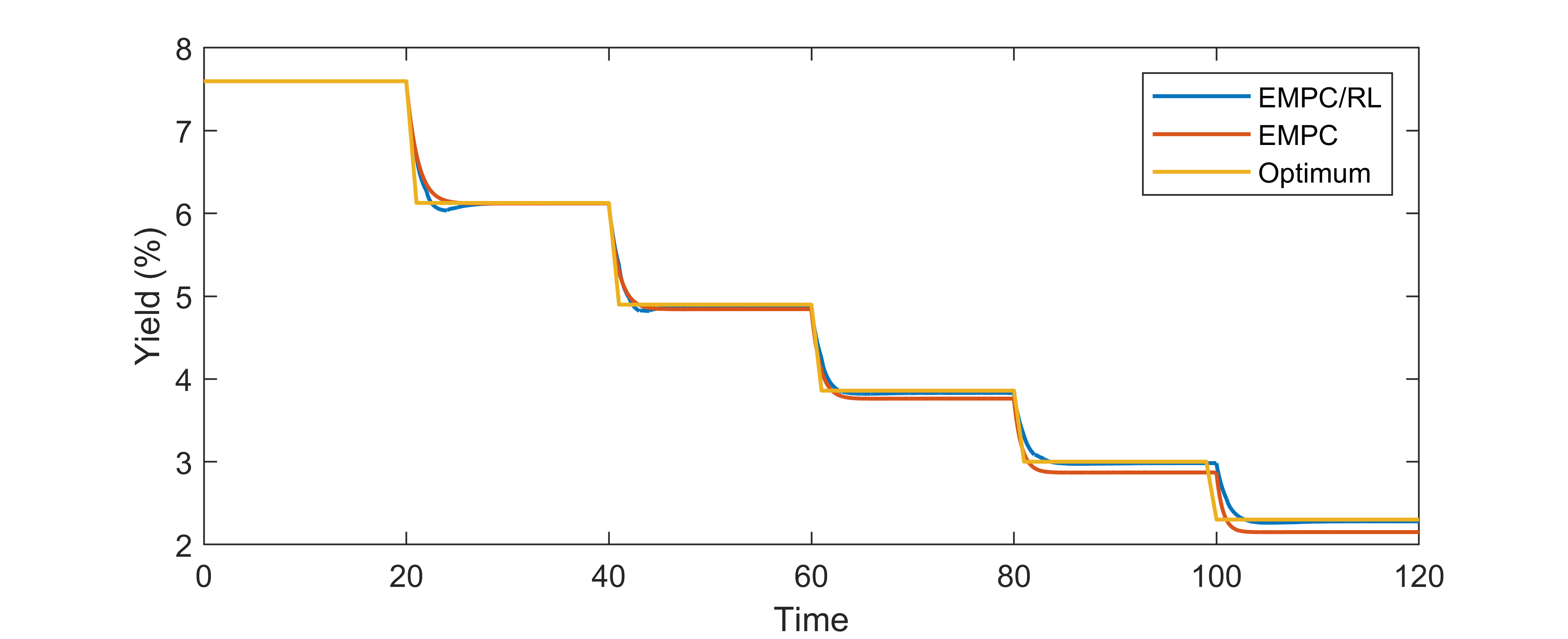}
\caption{Comparison of yield achieved by EMPC-RL framework and EMPC alone relative to optimal yield.}
\label{yield_all}
\end{figure}

Finally, Figure \ref{yield_all} compares the performance of using EMPC alone, and using the EMPC-RL scheme relative to the maximum yield that can be achieved. Because the catalyst experiences greater deactivation, and the model starts to deviate significantly, the yield achieved by EMPC alone becomes increasingly suboptimal. However, when EMPC-RL is applied, optimal yield is maintained throughout the cycle.

Table \ref{improvement_overEMPC} shows the yield improvement resulting from integrating EMPC and RL for each step. Note that the greater the model mismatch, the greater the improvement. For a mismatch of only $5\%$, the percent improvement over EMPC is $6.04\%$.

\begin{table}[H]
\centering
\caption[width=0.5\textwidth]{Calculation of the percentage of improvement as a result of using EMPC-RL relative to using EMPC alone}
\label{improvement_overEMPC}
\begin{tabular}{cc}
\hline
Step & \% Improvement over EMPC \\
\hline
0        & 0                                                                                                 \\
1       & 0                                                                                                \\
2       & 0.6                                                                                                 \\
3       & 2.55                                                                                                  \\
4       & 3.94                                                                                                  \\
5       & 6.04                                                                                             
\end{tabular}
\end{table}

\subsubsection{Performance with process noise}

\begin{figure}[H]
\centering
\includegraphics[width=0.8\textwidth]{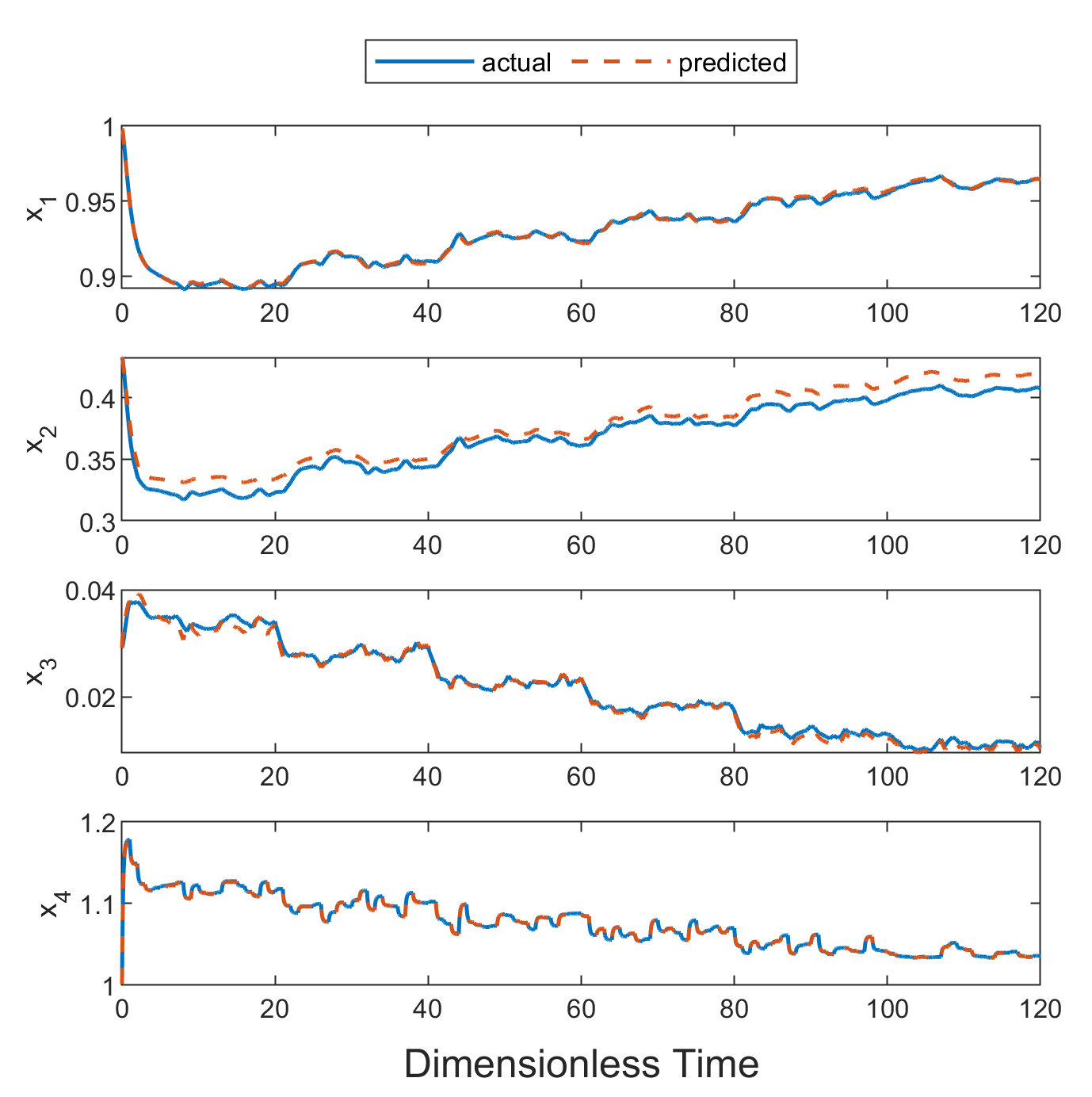}
\caption{State trajectories of process described by Eqs. \ref{ODEs} for EMPC-RL framework when process noise is added.}
\label{x_noise}
\end{figure}

To evaluate the robustness of the EMPC-RL scheme when there were process disturbances ($d(t)\neq0$), white Gaussian noise was introduced. As Figure \ref{x_noise} shows, the RL agent was able to track the observed states without causing any instabilities to the system.

\begin{figure}[H]
\centering
\includegraphics[width=1\textwidth]{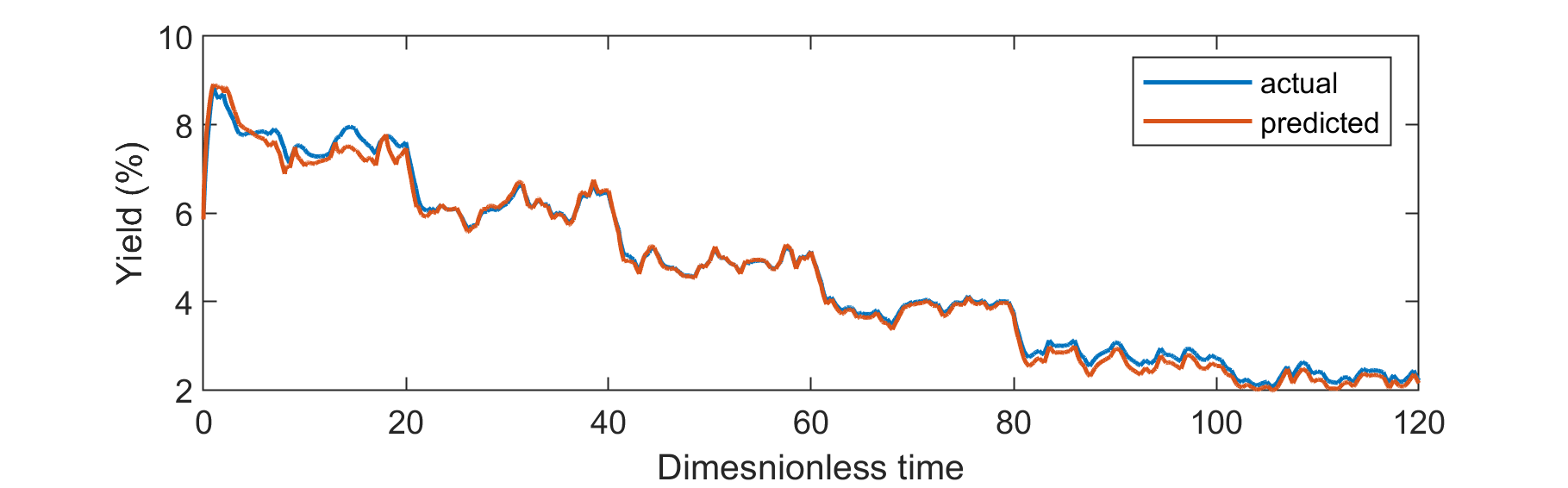}
\caption{Comparison of the predicted yield with the plant yield when the proposed framework is used under process noise}
\label{yield_noise}
\end{figure}

Figure \ref{yield_noise} presents the performance of the EMPC-RL scheme in terms of yield prediction when process noise is added.

\begin{figure}[H]
    \centering
    \includegraphics[width=1\textwidth]{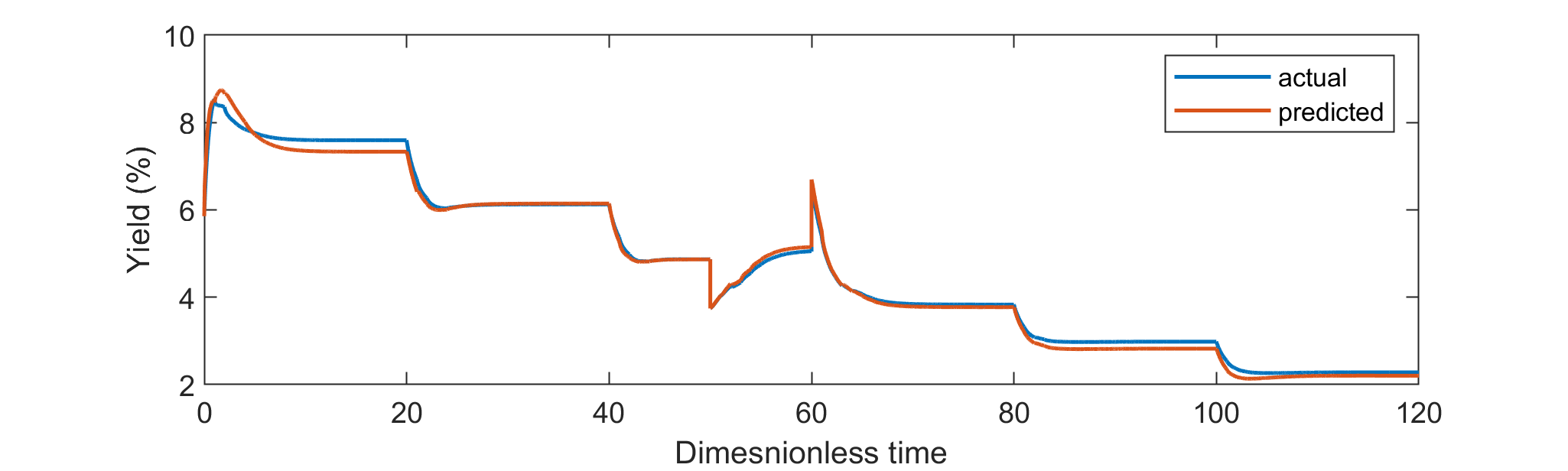}
    \caption{Comparison of predicted yield with  plant yield}
    \label{CE_spike}
    \end{figure}

The RL agent was not trained with noise or any other disturbances, yet it was able to correct the model since the RL agent was trained to learn a policy that corrects model-plant deviation, regardless of the cause. To further clarify this point, a $30\%$ spike in the inlet ethylene concentration from 50sec to 60sec was simulated. As can be seen in Figure \ref{CE_spike}, this RL-based EMPC scheme was robust to other process disturbances, as well as white Gaussian noise.

\subsubsection{Simulation results using three manipulated variables}

The EMPC formulated in Eqs. \ref{EMPC} included one manipulated variable, $u_3$. To further demonstrate the robustness of the RL agent and its ability to act in unfamiliar situations, the EMPC formulation was modified to include three manipulated variables: $u_1$, $u_2$, and $u_3$. The variable $u_3$ was constrained as stated earlier, while $u_2$ and $u_3$ were constrained as follows:
\begin{align*}
 0.071\le u_1\le 0.71 ~~~ \& ~~~ 0.25\le u_2\le 2.5
\end{align*}
Also, since $u_1$ and $u_2$ were no longer fixed during the length of operation, the stage cost was modified to be as follows:
$$l_e=\frac{\ x_3\ x_4}{u_2}$$

\begin{figure}[H]
\centering
\includegraphics[width=0.8\textwidth]{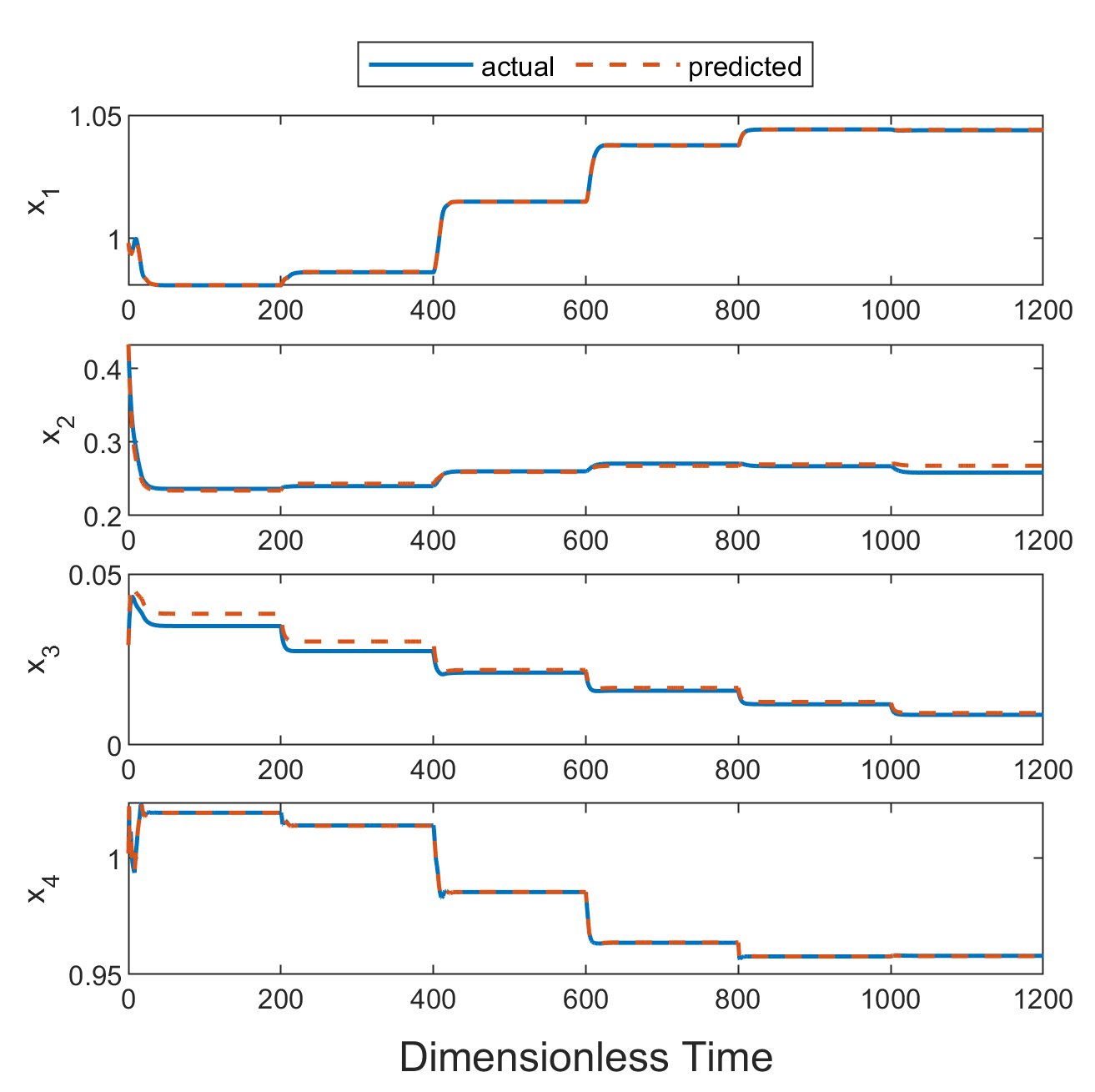}
\caption{State trajectories of process described by Eqs. \ref{ODEs} for EMPC-RL framework with three manipulated variables}
\label{x_3u}
\end{figure}

\begin{figure}[H]
\centering
\includegraphics[width=\textwidth]{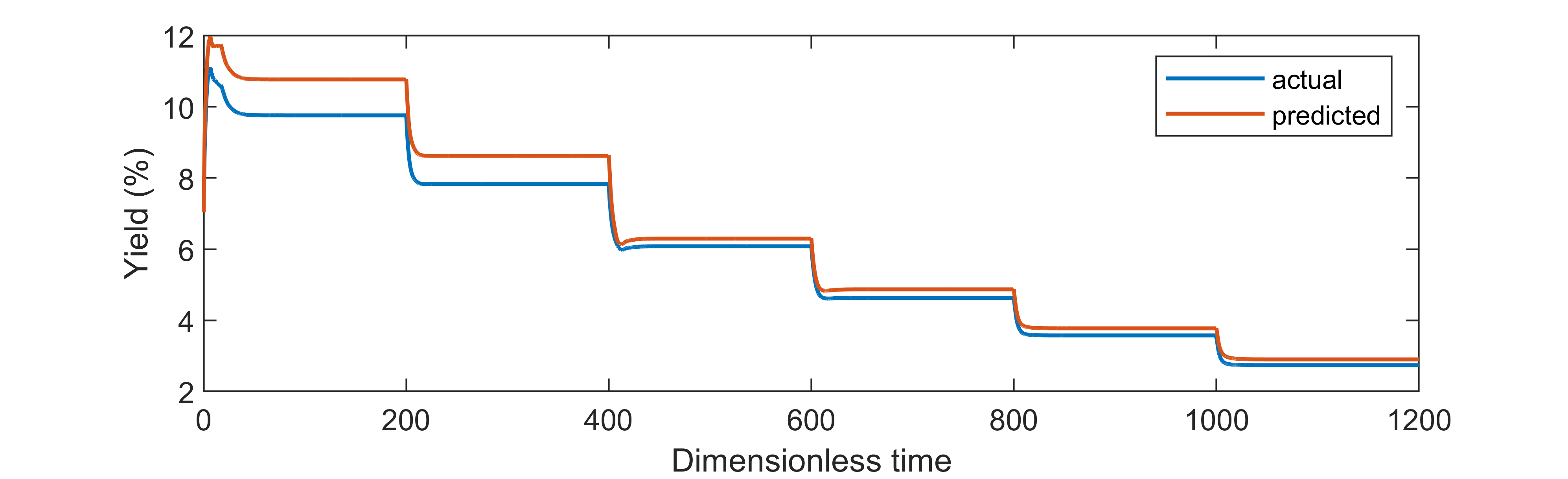}
\caption{Comparison of predicted yield with the plant yield when proposed framework was used with three manipulated variables}
\label{yield_3u}
\end{figure}

Figures \ref{x_3u} and \ref{yield_3u} show the performance of the EMPC-RL scheme when the EMPC was modified to include three manipulated variables, while using the same RL agent that was trained on the EMPC with one manipulated variable. It can be seen that the RL agent was able to track the observed state of the plant with an error of $10\%$ or less. This illustrates that the RL agent successfully learned a policy that determined how the model parameters should be modified in response to changes in the plant, such as catalyst deactivation, regardless of the controller.

\section{Conclusion}
This work presented a novel framework for integrating deep reinforcement learning with economic MPC for the purpose of operating chemical reactors at near optimal conditions in the presence of plant and model mismatch. The framework was applied to the oxidation of ethylene to ethylene oxide and demonstrated superior performance and improved yield in the desired product. It was also demonstrated that the RL agent can effectively act in environments that are different from those in which it was trained. Although policy learned by the RL agent to update the kinetic parameters was suboptimal, an improvement of yield performance was still achieved. As the field of reinforcement learning continues to progress, and more efficient algorithms are devised, better performance of this RL-EMPC framework is expected.

\section{Acknowledgments}
This work was funded by the KAUST Office of Sponsored Research (Grant OSR-2019-CRG7-4077). Fahad Albalawi acknowledges Taif University for their support via Taif University Researchers Supporting Project (TURSP-2020/97).
\bibliographystyle{elsarticle-num}
\bibliography{main.bib}
\end{document}